# Effects of wave interaction on ignition and deflagration-to-detonation transition in ethylene/air mixtures behind a reflected shock

Zhiwei Huang, Huangwei Zhang*

*Department of Mechanical Engineering, National University of Singapore, 9 Engineering Drive 1, Singapore 117576, Republic of Singapore*

**Abstract**

Dynamics of ethylene autoignition and Deflagration-to-Detonation Transition (DDT) in a one-dimensional shock tube are numerically investigated using a skeletal chemistry including 10 species and 10 reactions. Different combustion modes are investigated through considering various premixed gas equivalence ratios (0.2−2.0) and incident shock wave Mach numbers (1.8−3.2). Four ignition and DDT modes are observed from the studied cases, i.e., no ignition, deflagration combustion, detonation after reflected shock and deflagration behind the incident shock. For detonation development behind the reflected shock, three autoignition hot spots are formed. The first one occurs at the wall surface after the re-compression of the reflected shock and contact surface, which further develops to a reaction shock because of "the explosion in the explosion" regime. The other two are off the wall, respectively caused by the reflected shock / rarefaction wave interaction and reaction induction in the compressed mixture. The last hot spot develops to a reaction wave and couples with the reflected shock after a DDT process, which eventually leads to detonation combustion. For deflagration development behind the reflected shock, the wave interactions, wall surface autoignition hot spot as well as its induction of reaction shock are qualitatively similar to the mode of detonation after incident shock reflection, before the reflected shock / rarefaction wave collision point. However, only one hot spot is induced after the collision, which also develops to a reaction wave but cannot catch up with the reflected shock. For deflagration behind the incident shock, deflagration combustion is induced by the incident shock compression whereas detonation occurs after the shock reflection. The chemical timescale increases after the reflected shock / contact surface collision, whereas decreases behind the incident and reflected shocks, as well as after the reflected shock / rarefaction wave interaction. Therefore, mixture reactivity behind the reflected shock is weakened by contact surface, but is intensified by rarefaction wave. The multi-dimensionality characteristics including reflected shock / boundary layer interactions, reflected shock bifurcation, destabilization and detonation, are present in a two-dimensional configuration. Planar autoignition occurs because of reflected shock compression and detonation combustion is formed first in the central region due to the collision of reflected shock wave / reflected contact surface. The left- and right-bifurcations of the separation region in the wall boundary layer are then sequentially ignited.

* Corresponding author. Tel: +65 6516 2557; E-mail: huangwei.zhang@nus.edu.sg.

## 1. Introduction

Due to the simplicity of geometrical configuration and convenience in controlling the thermodynamic conditions of the post-shock gas, shock tube experiments are popularly used to measure ignition delay time of various fuels [1] or investigate two-phase gas – droplet interactions [2]. Two essential characteristics of the gas dynamics in shock tubes (either reactive or non-reactive) are various wave interactions (e.g., those between reflected shock, contact surface and rarefaction wave) [3–5] and shock-boundary layer interactions [6–8]. However, to prolong the test time usable in ignition delay measurements, "tailored" driver-driven gas interface is widely used in previous shock tube studies, experimentally [9–11], numerically [12–15] and analytically [16,17]. In such case (also termed as matched condition [17]), the wave interactions are avoided and the strength of reflected shock wave is unchanged when it transmits through the tailored contact surface.

Nevertheless, the interactions between the reflected shock wave and contact surface as well as rarefaction wave may be important for the practical ignition behaviors of combustible gas [5,18]. One-dimensional (1D) interaction of a detonation wave with a contact discontinuity is investigated analytically and experimentally for oxygen-hydrogen mixture [5]. The results show that the shock can either be amplified or attenuated when it transmits through the contact surface and then propagates into an inert gas (helium/air mixture), depending on the reflection type at the contact surface as well as the ratio of acoustic impedance across it. The reflection type is found to depend on the ratio of internal energies across the contact surface [19]. However, the ignition and combustion development of hydrocarbon fuels like ethylene, under complex wave interactions (include but not limit to shock / contact surface interaction), are still not fully understood from the above studies. Furthermore, the behavior of the shock after its interaction with contact surface, e.g., when it further propagates into a combustible medium, has not been studied.

Generally, two ignition modes, i.e., mild [16,20,21] and strong (or sharp) [22–24] ignitions are observed both experimentally and numerically. According to Meyer *et al.* [24], mild ignition starts from distinct flame kernels which grow slowly, whereas strong ignition is planar ignition that covers the cross-section of a shock tube instantaneously. In the work of Yamashita *et al.* [6], the ignition modes are classified as near-wall (strong) and far-wall (mild) ignitions. The local ignition hot spots in mild ignition are induced by the non-uniformity behind reflected shock waves [6], which can occur in e.g., the boundary region near the side-wall of the shock tube [25–27], the bulk flow [28,29], or the both regions but with some delay in time [16,30]. The switch between the different localized ignition modes is found to be sensitive to e.g., ignition temperature [6,31,32] and shock wave intensity [8,33,34], which implies that different mechanisms may be responsible for the formation of ignition hot spot. Therefore, it is of interest to further investigate ignition and deflagration / detonation development under complex wave interactions for hydrocarbon fuels.

In this work, detailed numerical studies are performed to investigate the autoignition and Deflagration-to-Detonation Transition (DDT) in ethylene/air mixtures subject to the shock wave reflection on the wall. A skeletal ethylene mechanism is used, which includes 10 species and 10 reactions [35]. Different premixed gas equivalence ratios and incident shock wave Mach numbers are studied. 1D and two-dimensional (2D) domains are used with high mesh resolutions to capture the detailed and unsteady gas dynamics.

The novelties of this study are two-fold. Firstly, the multi-wave systems are considered, e.g., the incident shock wave, contact surface and rarefaction wave as well as their reflected counterparts. This differs from the previous work, e.g., Refs. [12,13,36], which deliberately avoid the wave interactions and therefore their effects on combustion evolutions cannot not be reproduced. Secondly, automated reaction analysis based on Chemical Explosive Mode Analysis (CEMA) [37–40] is performed through an eigen-decomposition of the chemical Jacobian matrix [38,41]. This enables us to extract the quantitative chemical information behind the wave−chemistry interactions, which is unveiled in the previous studies.

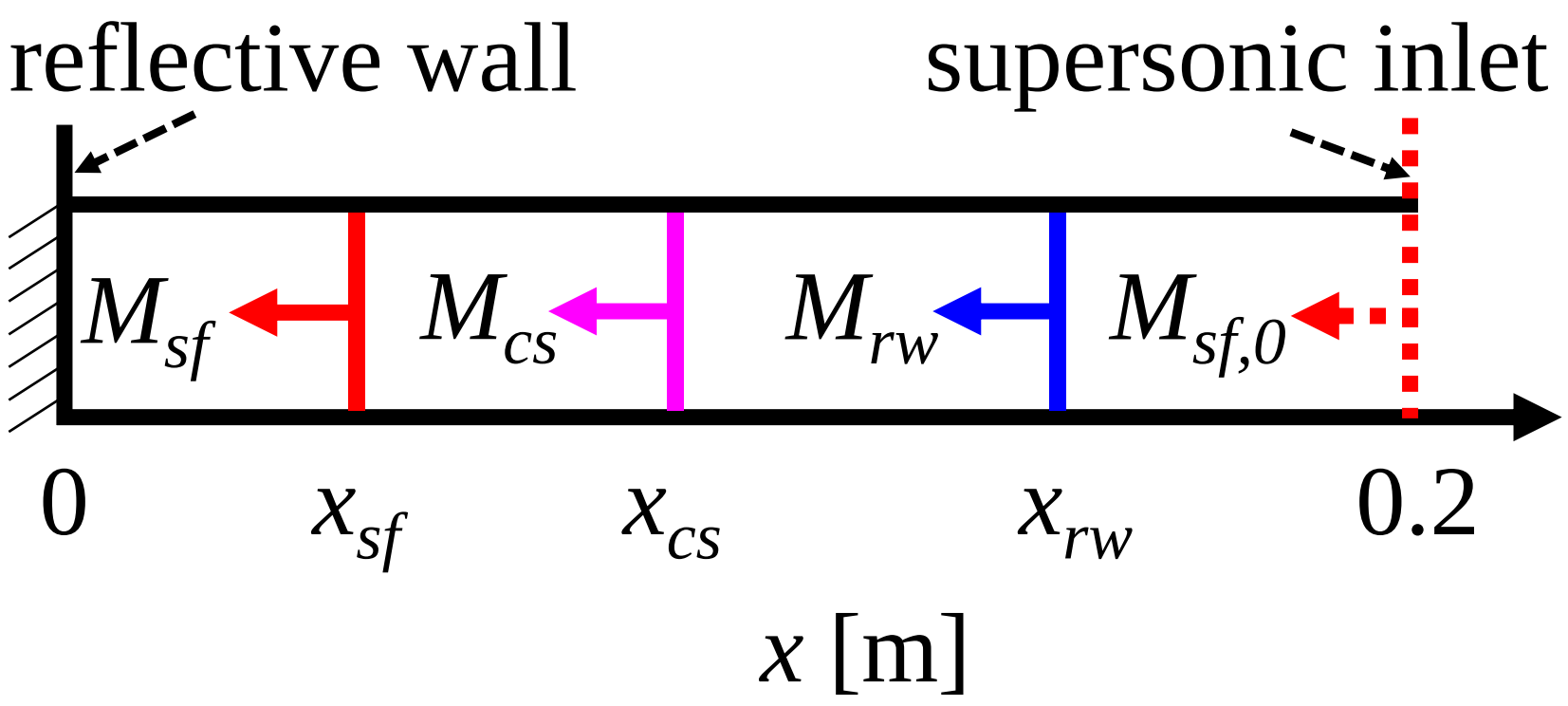

**Fig. 1.** Computational domain for a one-dimensional shock tube.

## 2. Physical problem

Autoignition and deflagration-to-detonation transition in combustible mixtures behind reflected shock waves with a semi-closed shock tube are studied. Figure 1 shows a schematic of the 1D computational domain, which starts at $x$ = 0 and is 0.2 m long ($x$-direction). The left boundary is a reflective, adiabatic non-slip wall, whereas the right is a supersonic inlet with Dirichlet conditions enforced for all variables. A left-propagating incident shock enters the domain at the right end ($x$ = 0.2 m) at $t$ = 0, with an initial Mach number $M_{sf,0}$. The leading shock is followed by a contact surface and a rarefaction wave, and their locations are marked as $x_{sf}$, $x_{cs}$ and $x_{rw}$ in Fig. 1. Their corresponding Mach numbers are $M_{sf}$, $M_{cs}$ and $M_{rw}$, respectively. The interactions between the left wall and incident shock wave would lead to a right-propagating reflected shock wave and therefore second compression of the combustible mixture. Note that the multi-dimensionality effects in experimental shock tubes may be also important because of boundary layer growth and/or shock / boundary layer interactions, which further leads to non-uniform ignition [6,7,42]. This will be examined through high-resolution two-dimensional simulations in Section 4.3.

The initial gas composition in the shock tube is ethylene/air mixture with equivalence ratios of $\phi_0$ = 0.2−2.0. The initial pressure and temperature are respectively 10 kPa and 300 K. The incident shock Mach numbers considered in this study are $M_{rw,0}$ = 1.8, 2.0, 2.4, 2.8 and 3.2. The gas conditions at the inlet, behind the incident shock, and behind the reflected shock for stoichiometric ethylene/air mixture (mass ratio $Y_{C2H4}/Y_{O2}/Y_{N2}$ = 0.064/0.218/0.718) are detailed in Table 1. Here, the studied conditions behind the reflected shock wave are comparable to the ethylene experiments of Penyazkov *et al.* ($p_{rs,0}$ = 5.9-16.5 atm, $T_{rs,0}$ = 1,060-1,520 K) [9], Saxena *et al.* ($p_{rs,0}$ = 2, 10, 18 atm, $T_{rs,0}$ = 1,000-1,650 K) [10], and Wan *et al.* ($p_{rs,0}$ = 0.97-20.54 atm, $T_{rs,0}$ = 721-1,307 K) [11].

**Table 1.** Gas conditions corresponding to different inflow Mach numbers in stoichiometric ethylene/air mixture. $p_{rw,0}$ / $T_{rw,0}$, $p_{sf,0}$ / $T_{sf,0}$ and $p_{rs,0}$ / $T_{rs,0}$ are respectively the pressures / temperatures at the inlet ($x$ = 0.2 m), behind incident shock and reflected shock.

| $M_{rw,0}$ | $p_{rw,0}$ [kPa] | $T_{rw,0}$ [K] | $p_{sf,0}$ [kPa] | $T_{sf,0}$ [K] | $p_{rs,0}$ [kPa] | $T_{rs,0}$ [K] |
|---|---|---|---|---|---|---|
| 1.8 | 56.6 | 472.4 | 79.6 | 646.9 | 364.1 | 1009.8 |
| 2.0 | 77.4 | 509.2 | 104.6 | 751.7 | 540.9 | 1215.4 |
| 2.4 | 147.5 | 590.5 | 177.1 | 1036.8 | 1122.5 | 1774.7 |
| 2.8 | 285.2 | 680.2 | 289.7 | 1443.2 | 2144.1 | 2588.9 |
| 3.2 | 554.6 | 777.5 | 456.2 | 2009.1 | 3797.0 | 3743.1 |

## 3. Numerical method and computational diagnostic tool

### *3.1. Numerical method*

The governing equations of mass, momentum, energy, and species mass fractions are solved for compressible, multi-species, reacting flows. The governing equations are solved by a density-based multi-component reactive flow solver, *RYrhoCentralFoam* [43–45]. It is developed from *rhoCentralFoam* solver in OpenFOAM 5.0 package [46]. The *RYrhoCentralFoam* solver is verified and validated with a series of benchmark cases against analytical solutions and/or experimental data in our recent work [45,47]. It can accurately predict gaseous detonation properties in different fuels (e.g., hydrogen and methane), including reaction-shock interaction, propagation speed, frontal structures, and cell size [45]. Recently, it is successfully used for simulations of supersonic combustion and detonations [44,45,47].

A second-order implicit Crank−Nicolson scheme [48] is used for temporal discretization. A Godunov-type Riemann-solver-free scheme developed by Kurganov *et al.* [49] is used, with Minmod flux limiter [50] for convective fluxes for accurate shock capturing. The diffusive fluxes are predicted with a second-order central differencing scheme [51]. The physical time step is $10^{-9}$−$10^{-10}$ s. Detailed information about the equations and numerical method can be found in Refs. [43–47,52].

The stiff ODE solver *seulex* [53] is used to integrate the chemical reaction system. A skeletal chemical mechanism for ethylene combustion is used, which contains 10 species ($C_2H_4$, $O_2$, CO, $H_2$, O, $CO_2$, OH, H, $H_2O$ and $N_2$) and 10 reactions [35]. The Arrhenius kinetic parameters can be found in Ref. [35], whilst the thermodynamic ones are estimated with JANAF polynomials [54]. This mechanism is validated against the detailed mechanism [55] (25 species and 77 reactions) over a range of operating conditions and the results agree well with the measured data for, e.g., evolutions of temperature, pressure and key species concentrations. It is used for simulations of supersonic flames [56,57] and detonations [58,59].

Three uniform meshes with 10,000, 20,000 and 40,000 cells are adopted to discretize the 1D domain in Fig. 1. They correspond to the cell sizes of 20, 10 and 5 μm, respectively. The grid dependence analysis is provided in the supplementary material, and it is found that the differences in autoignition and DDT development with three meshes are negligible. Therefore, in the subsequent analysis, the resolution of 10 μm is used. This leads to about 98 cells in the half-reaction zone of Chapman–Jouguet (C−J) detonation under the conditions behind the reflected shock wave with $M_{rw,0}$ = 2.0 (see Table 1).

### 3.2. Chemical reaction analysis

CEMA [37–40] is an automated approach for studying ignition and detonation development in premixed gas. It can extract the comprehensive reaction information from local chemical Jacobian matrix [38,41]. For a reaction system, the evolutions of thermochemical composition follow

$$\frac{D\boldsymbol{\varphi}}{Dt} = \boldsymbol{g}_{\omega}(\boldsymbol{\varphi}) + \boldsymbol{s}(\boldsymbol{\varphi}), \tag{1}$$

where $\boldsymbol{\varphi} = (c_1, c_2, \ldots, c_N, T)$ is the vector of species molar concentrations $c_i$ and temperature $T$ from the detailed reactive flow simulations. $D(\cdot)/Dt$ is the material derivative. The term $\boldsymbol{g}_{\omega}(\boldsymbol{\varphi})$ is the vector of chemical source terms, whereas $\boldsymbol{s}(\boldsymbol{\varphi})$ denotes all the non-chemical terms. In CEMA method, eigen-analysis of the local chemical Jacobian matrix is performed [38,39,41], i.e.,

$$\frac{D\boldsymbol{g}_{\omega}(\boldsymbol{\varphi})}{Dt} = J \cdot \frac{D\boldsymbol{\varphi}}{Dt} = J \cdot \boldsymbol{g}_{\omega}(\boldsymbol{\varphi}), \tag{2}$$

where $J = \frac{\partial \boldsymbol{g}_{\omega}(\boldsymbol{\varphi})}{\partial \boldsymbol{\varphi}}$ is the local chemical Jacobian matrix. Chemical modes are associated with the eigenvalues of $J$, among which the one with the maximum real part is denoted as $\lambda_e$. A chemical explosive mode is identified when the real part of $\lambda_e$, $Re(\lambda_e)$, is positive.

The contribution of species or temperature to a CEM is quantified through explosion index [38,41], i.e.,

$$EI_j = max\{sign[Re(\lambda_e)], 0\} \cdot \frac{diag|\boldsymbol{l}_e \boldsymbol{r}_e|_j}{\sum_{i=1}^{N+1} diag|\boldsymbol{l}_e \boldsymbol{r}_e|_i}, \tag{3}$$

where $sign(x)$ and $max(x,y)$ are respectively the sign and maximum functions, $diag|\cdot|_j$ is the absolute value of diagonal element for $j$-th variable, $\boldsymbol{l_e}$ and $\boldsymbol{r_e}$ are the left and right eigenvectors corresponding to $\lambda_e$. $EI_j$ closes to 1 means that the corresponding variable (species concentration or temperature) is dominant in the CEM. Full details of the CEMA method can be found in Refs. [38,39,41].

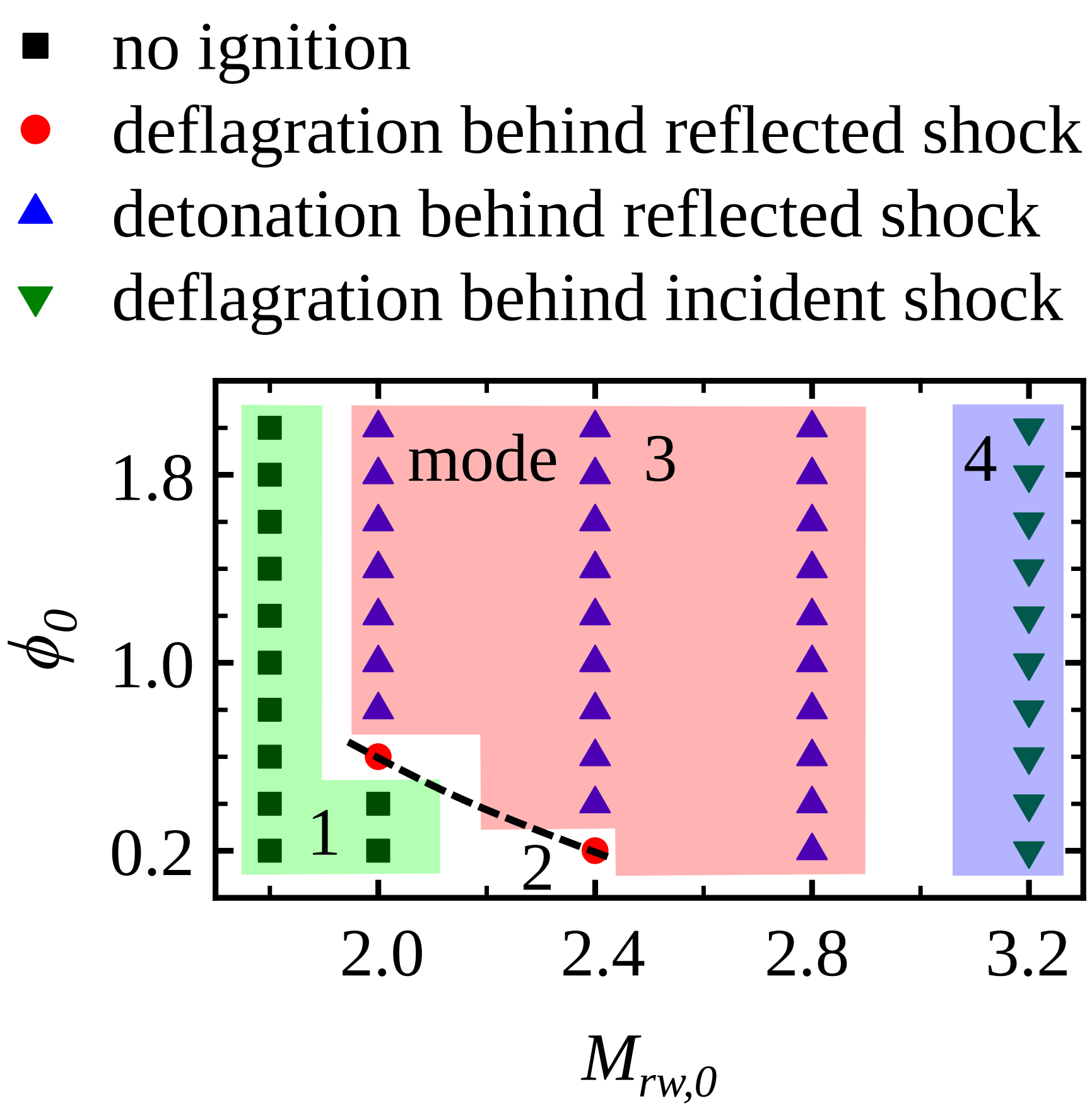

**Fig. 2.** Diagram of the combustion modes in ethylene/air mixture subject to incident/reflected shocks. The numbers indicate four combustion modes.

## 4. Results and discussion

### *4.1. Combustion mode*

Four combustion modes of ethylene/air mixtures subject to incident / reflected shock waves are observed based on our simulation results: (1) no ignition, (2) deflagration combustion behind reflected shock, (3) detonation combustion behind reflected shock, and (4) deflagration combustion behind the incident shock wave (also develops to detonation after the shock is reflected at the wall). A combustion mode map is shown in Fig. 2, which is parameterized by premixture equivalence ratio and inflow Mach number. It is found that under low inflow Mach numbers and/or low equivalence ratios, modes 1 (e.g., $M_{rw,0}$ ≤1.8, or $M_{rw,0}$ = 2.0 but $\phi_0 \leq 0.4$) and 2 (two cases with $M_{rw,0}$ = 2.0 and $\phi_0$ = 0.6, $M_{rw,0}$ = 2.4 and $\phi_0$ = 0.2) are more likely to occur. Mode 3 becomes more prevalent when the inflow Mach number is further increased. When $M_{rw,0}$ = 3.2, mode 4 occurs for all the considered equivalence ratios. One can also see that the dependence of combustion mode on the mixture equivalence ratio is weaker than that of inflow Mach number. For instance, when $\phi_0$ is above 0.8, the predicted modes in Fig. 2 are solely affected by the $M_{rw,0}$. In the following, detailed transients in combustion modes 2−4 and the underlying interactions between chemical reaction and gas dynamics will be discussed through the representative cases.

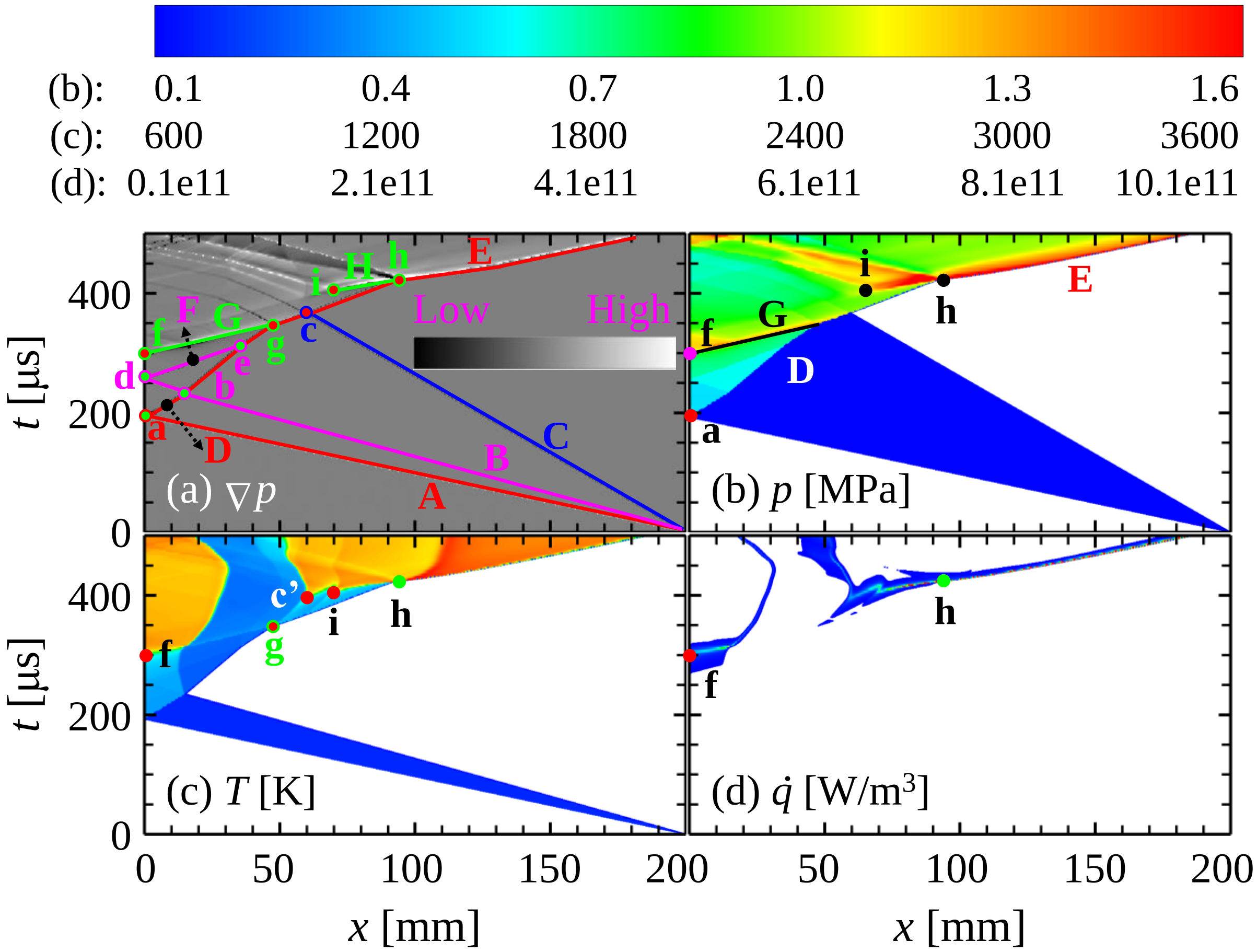

**Fig. 3.** *x-t* diagrams of (a) pressure gradient, (b) pressure, (c) temperature and (d) heat release rate. $M_{rw,0}$ = 2.0 and $\phi_0$ = 1.0.

*4.1.1. Autoignition and detonation development*

Figure 3 shows the evolutions of pressure gradient, pressure, temperature, and heat release rate in *x-t* diagram. The initial conditions are $M_{rw,0} = 2.0$ and $\phi_0 = 1.0$, which corresponds to mode 3. In Fig. 3(a), the incident shock wave A impinges on the wall at point a ($t \approx 193$ μs) and is reflected (shock wave D). The latter intersects with incident contact surface B at point b ($x \approx 16.2$ mm and $t \approx 239$ μs), and both are decelerated after they collide because their velocities are in opposite directions. Then the contact surface B impinges on the wall at point d ($t \approx 255$ μs), and a reflected contact surface F is formed. It subsequently merges with the reflected shock D at point e ($x \approx 35.8$ mm and $t \approx 314$ μs) and D is intensified because the pressure gradients of F and D are in the same direction.

Moreover, a shock wave G is induced at point f ($t \approx 302$ μs). This is caused by "the explosion in the explosion" [9,13,60], i.e., the near-wall autoignition hot spot f induced by the shock compression, manifested by locally high pressure, temperature, and heat release rate in Figs. 3(b)-3(d), respectively. This reaction shock G propagates towards the shocked gas, and then catches up with the reflected shock D at point g ($x \approx 47.4$ mm and $t \approx 348$ μs) and part of it is reflected back (see the backward propagating bifurcation at point g). Nevertheless, the reflected shock D is slightly intensified after point g (flatter slope of the trajectory), which then collides with the incident rarefaction wave C at point c ($x \approx 59.8$ mm and $t \approx 366$ μs). Both waves are intensified, and a second reactive hot spot c' is initiated, which is originated from D-C collision location c.

A third reactive hot spot appears in the compressed mixture at point i ($x \approx 70.2$ mm and $t \approx 406$ μs), which generates two bifurcated, right- and left-running, reaction waves. The former reaction wave, H, catches up with D at point h ($x \approx 93.0$ mm and $t \approx 423$ μs) after a DDT process. D and H mutually enhance and couple with each other, thereby generating a new detonation wave E at point h. The propagating speed of detonation wave E is 1,787 m/s, which is close to the C–J speed (1,797 m/s based on the gas conditions behind the incident shock, see Table 1).

Figure 4 shows the chemical explosive mode and explosion indices (Eq. 3) of temperature and dominant species (i.e., OH and O radicals) in $x$-$t$ diagrams. The CEM is visualized through

$$\lambda_{cem} = max\{sign[Re(\lambda_e)], 0\} \cdot log_{10}[1 + |Re(\lambda_e)|]. \quad (4)$$

In Fig. 4(a), only the mixture roughly between the incident shock A and contact surface B is chemically explosive before the shock reflection on the wall. However, $\lambda_{cem}$ is relatively low there, which indicates that the chemical timescales are large (say, above 1 s). After the reflected shock wave, the near-wall mixture (before point b) is found to have increased propensity of chemical explosion (high $\lambda_{cem}$), sequentially experiencing chemical runaway (or chain-branching reactions), thermal runaway, and ultimate autoignition (i.e., point f), as can be seen from Figs. 4(b) to 4(d). Note that chemical (thermal) runaway corresponds to a high dependence on radical species (temperature). It is also seen that on the left side of point b', across the reaction shock G the chemical mode abruptly transitions from explosive one to dissipative one (white area, i.e., the slowest decaying mode in a non-explosive mixture [38,39]). Moreover, a significant increase in $\lambda_{cem}$ is observed behind G, on the right side of point b', with high chemical runaway and hence strong chain-branching reactions (see explosion indices for OH and O in Figs. 4c and 4d). As time increases, thermal runaway propensity becomes significant, ultimately generating the second and third hot spots, c' and i (see Figs. 3b and 3c). When detonation is developed, the CEM region is only observed in the induction zone of the detonation wave E, which is consistent with the observations from two-dimensional detonation structures [61].

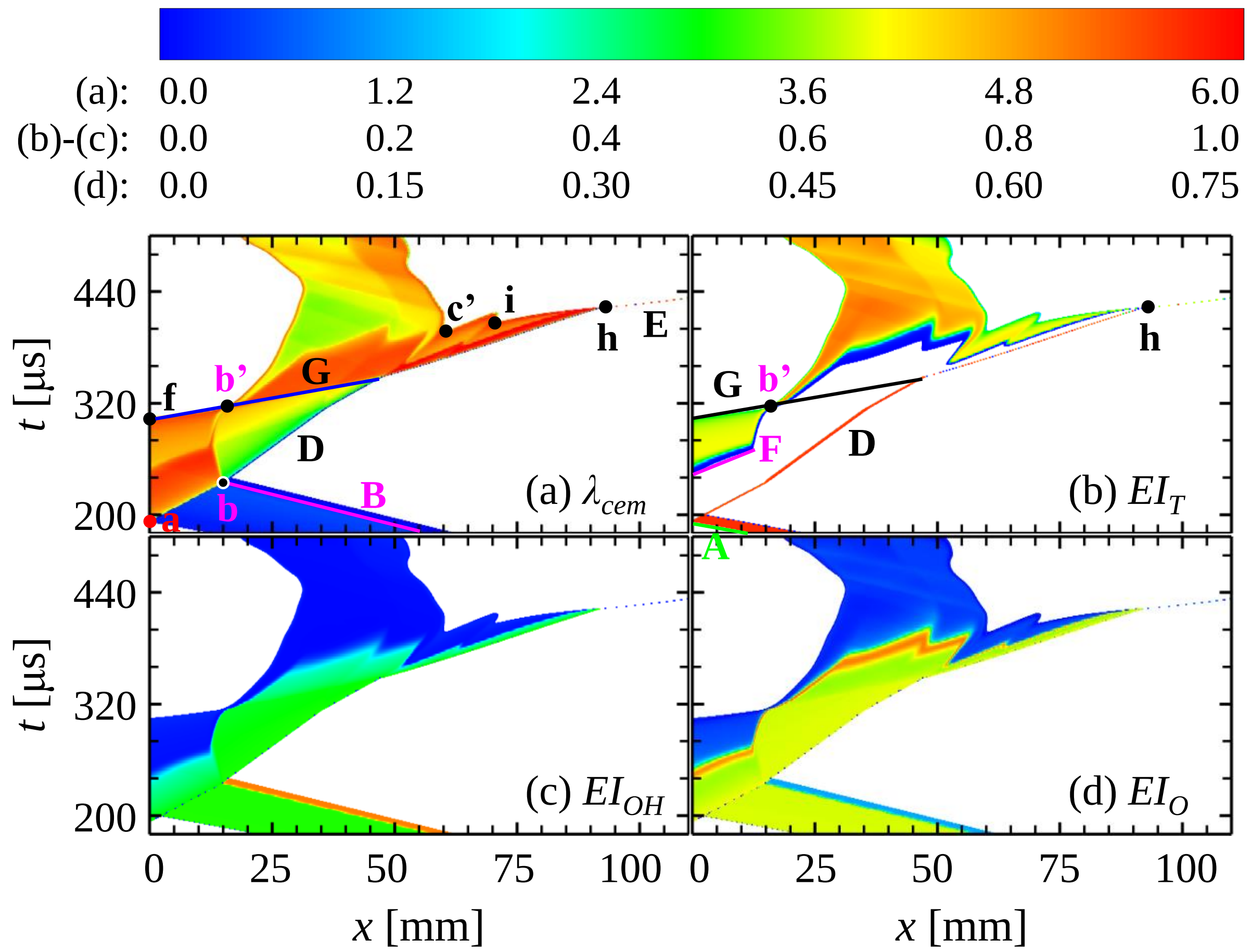

**Fig. 4.** *x-t* diagrams of (a) CEM distribution and explosion indices for (b) temperature, (c) OH and (d) O radicals.

To further elaborate on the interactions between the gas dynamics and chemical reactions, Fig. 5(a) shows the evolutions of chemical timescale, $t_e$, (i.e., the reciprocal of $|Re(\lambda_e)|$ from CEM [38]) at $x = 0$, 47.4, 70.2 and 93.0 mm, which respectively correspond to the locations of points a, g, i and h in Fig. 3(a). Figure 5(b) shows the evolutions of OH mass fraction $Y_{OH}$ at these locations. In Fig. 5(a), the mixture is not chemical explosive (e.g., $t_e > 10^5$ s) before the incident shock wave arrives. When the incident shock wave sweeps the four locations at $t \approx 193$, 148, 126 and 104 μs, respectively, there is a sharp decrease of $t_e$, to about 1 s (indicated by A1). However, $t_e$ recovers to above $10^3$ s after the incident contact surface passage (at $t \approx 203$, 173, and 143 μs) for the latter three points. For the wall surface $x = 0$, it is re-compressed by the reflected shock wave, and $t_e$ decreases to about $10^{-5}$ s (i.e., after point a). Then $t_e$ slightly increases due to thermal expansion from combustion heat release (which leads to decreased wall pressure). This ends at point d, when the incident contact surface is reflected on the wall. The mixture is compressed again and therefore $t_e$ decreases. Meanwhile, the OH radical rapidly increases after d and peaks at f, where finally the first autoignitive spot is formed as seen in Fig. 3(a). After that slow recombination reactions proceed and $t_e$ nearly keeps constant whereas OH is slowly reacted.

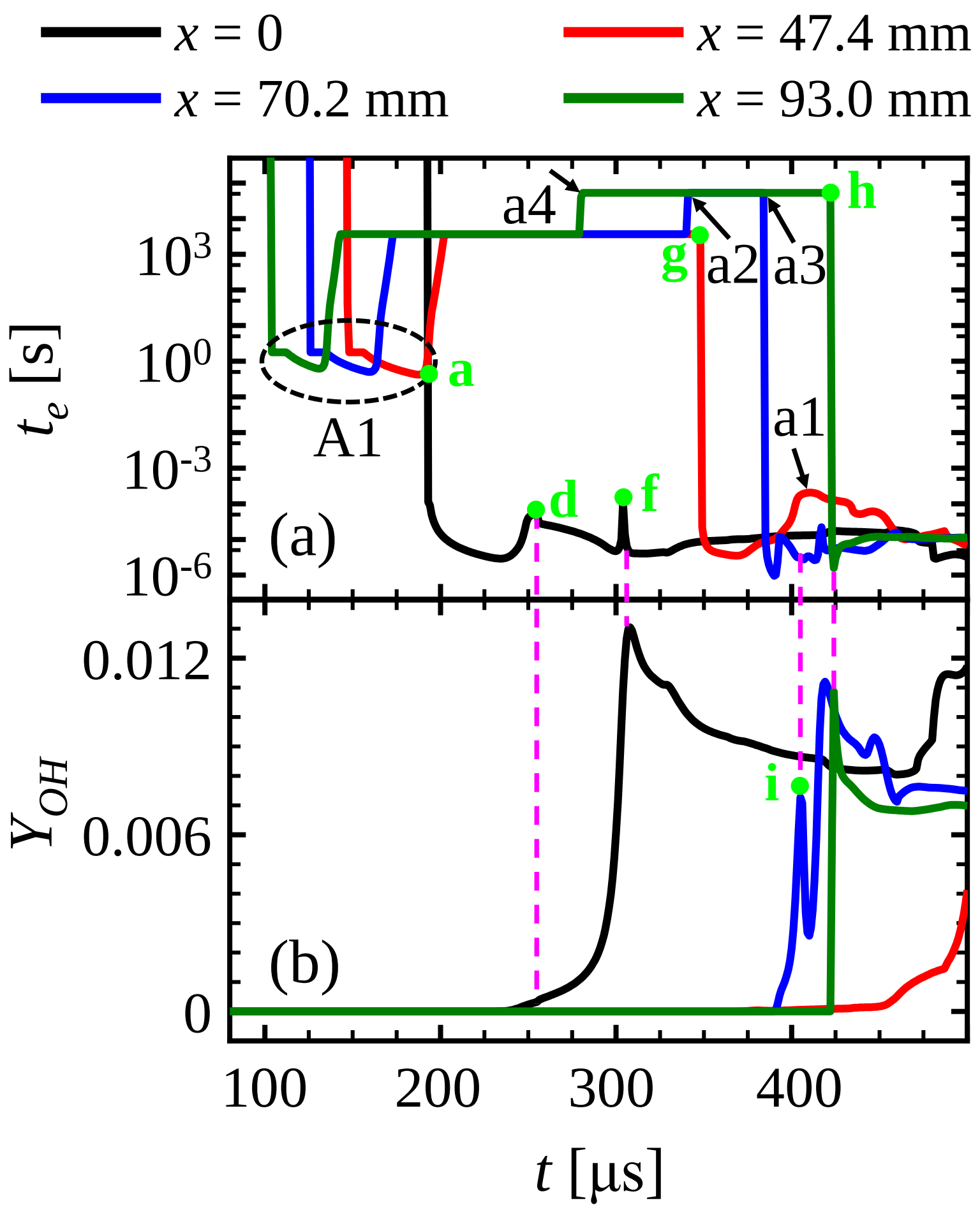

**Fig. 5.** Evolutions of (a) chemical timescale (in logarithmic scale) and (b) OH mass fraction at various locations. Letter symbols same as in Figs. 3 and 4.

For $x = 47.4$ mm, $t_e$ sharply decreases from above $10^3$ s (i.e., the state behind the incident contact surface), to about $10^{-5}$ s (the state behind G in Fig. 3a) at point g. Then $t_e$ slowly decreases because of continuous chain-branching reactions, but it peaks at $t \approx 409$ μs (a1) because the combustion is weakened by the incident rarefaction wave. After that, $t_e$ slowly decreases, whereas OH radical is slowly built up, indicating lasting reaction induction at this location.

For the reactive hot spot at $x = 70.2$ mm, the chemical timescale $t_e$ increases to above $10^4$ s behind the incident rarefaction wave at $t \approx 343$ μs (a2), which ends when the reflected shock wave D arrives here at $t \approx 384$ μs (a3). Meanwhile, pronounced OH radical occurs since then, which peaks when the autoignitive spot occurs at point i. There is another peak of $Y_{OH}$ after point i, which is caused by the left-propagating reaction wave originated from point h (see Fig. 3a). For the detonation initiation location at $x = 93.0$ mm, it also experiences the influences of incident shock wave and contact surface, as well as incident rarefaction wave ($t \approx 281$ μs, a4 in Fig. 5a). Then $t_e$ decreases to about $10^{-6}$ s while $Y_{OH}$ sharply increases at point h, when the detonation flame front is formed.

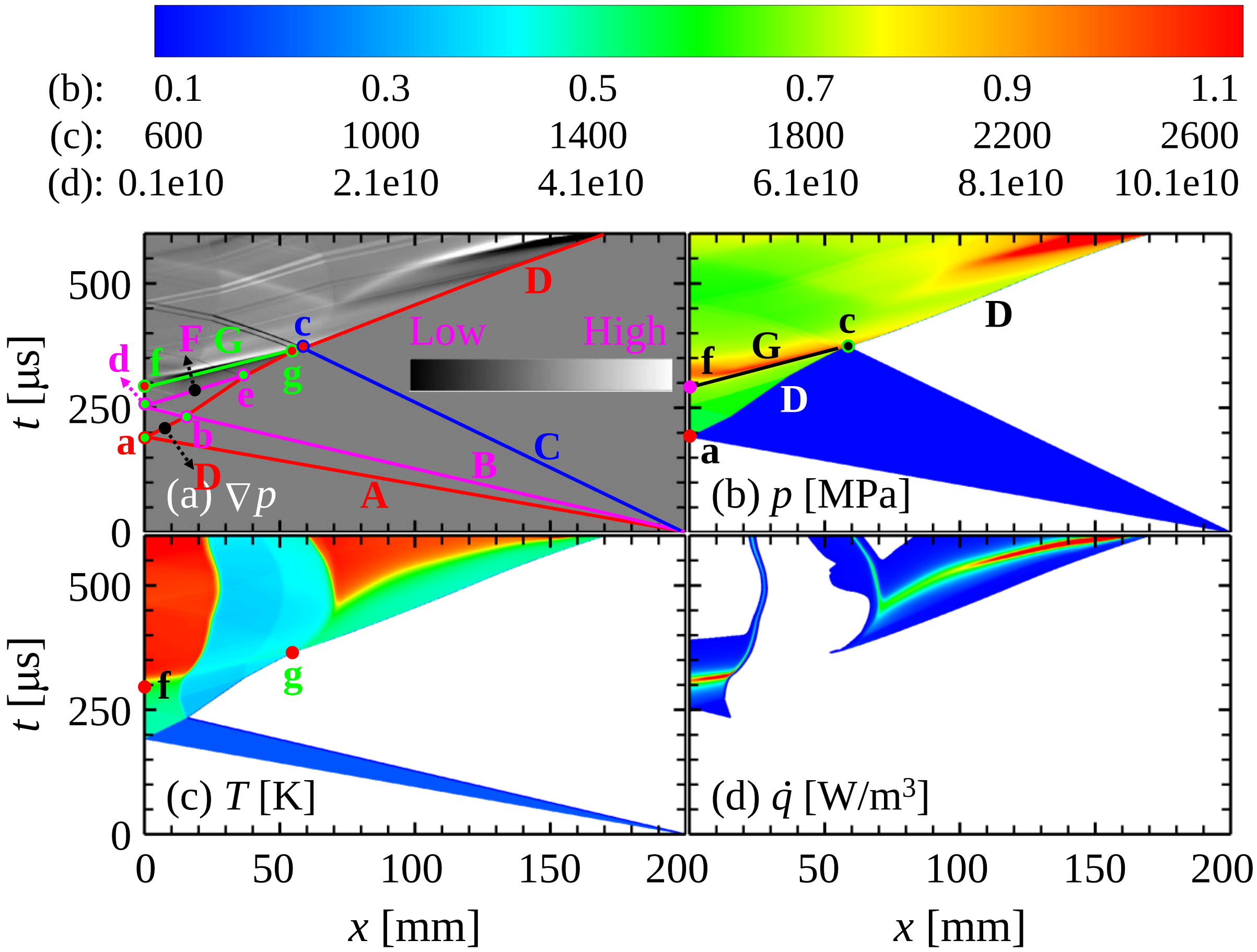

**Fig. 6.** *x-t* diagrams of (a) pressure gradient, (b) pressure, (c) temperature and (d) heat release rate. $M_{rw,0} = 2.0$ and $\phi_0 = 0.6$.

*4.1.2. Deflagration flame propagation*

Figure 6 shows the evolutions of pressure gradient, pressure, temperature, and heat release rate in *x-t* diagram. The initial conditions are $M_{rw,0}$ = 2.0 and $\phi_0$ = 0.6, which corresponds to mode 2. Different from the stoichiometric results in Section 4.1.1, in this case, only deflagration combustion is developed after the reflected shock wave. The interactions between the leading shock, contact surface, rarefaction wave and their reflections on the wall, as well as the reaction shock wave G before point c are qualitatively similar to those discussed in Section 4.1.1. After point c, only one hot spot is present, which can be seen from Fig. 6(c). The right-running reaction wave cannot catch up with the leading shock to support a propagating detonation, before the latter arrives at the right end. In Fig. 6(a), it is seen that the reaction front (the white zone following D) significantly lags behind D. With further decreased equivalence ratio (e.g., $\phi_0$ = 0.2), there is even no propagating reaction front behind the leading shock (off the wall, no hot spot can be formed), and only the near-wall mixture burns after shock compression.

Figure 7 shows the chemical explosive mode and explosion indices of temperature and dominant species in *x-t* diagram. In Figs. 7(a)-7(d), the evolutions of $\lambda_{cem}$, $EI_T$, $EI_{OH}$ and $EI_O$ are qualitatively similar to the counterparts in Fig. 4 before point c. Beyond point c, the CEM regions exist between the leading shock and reaction wave (see Fig. 7a). Within these regions, there are two stages, i.e., chemical runaway immediately behind the shock wave and thermal runaway ahead of the reaction wave. They can be observed from the local explosion indices of species and temperature, respectively, in Figs. 7(b)-7(d). The reaction front has a propensity to catch up with the leading shock D. However, they are still not fully coupled within the computational domain.

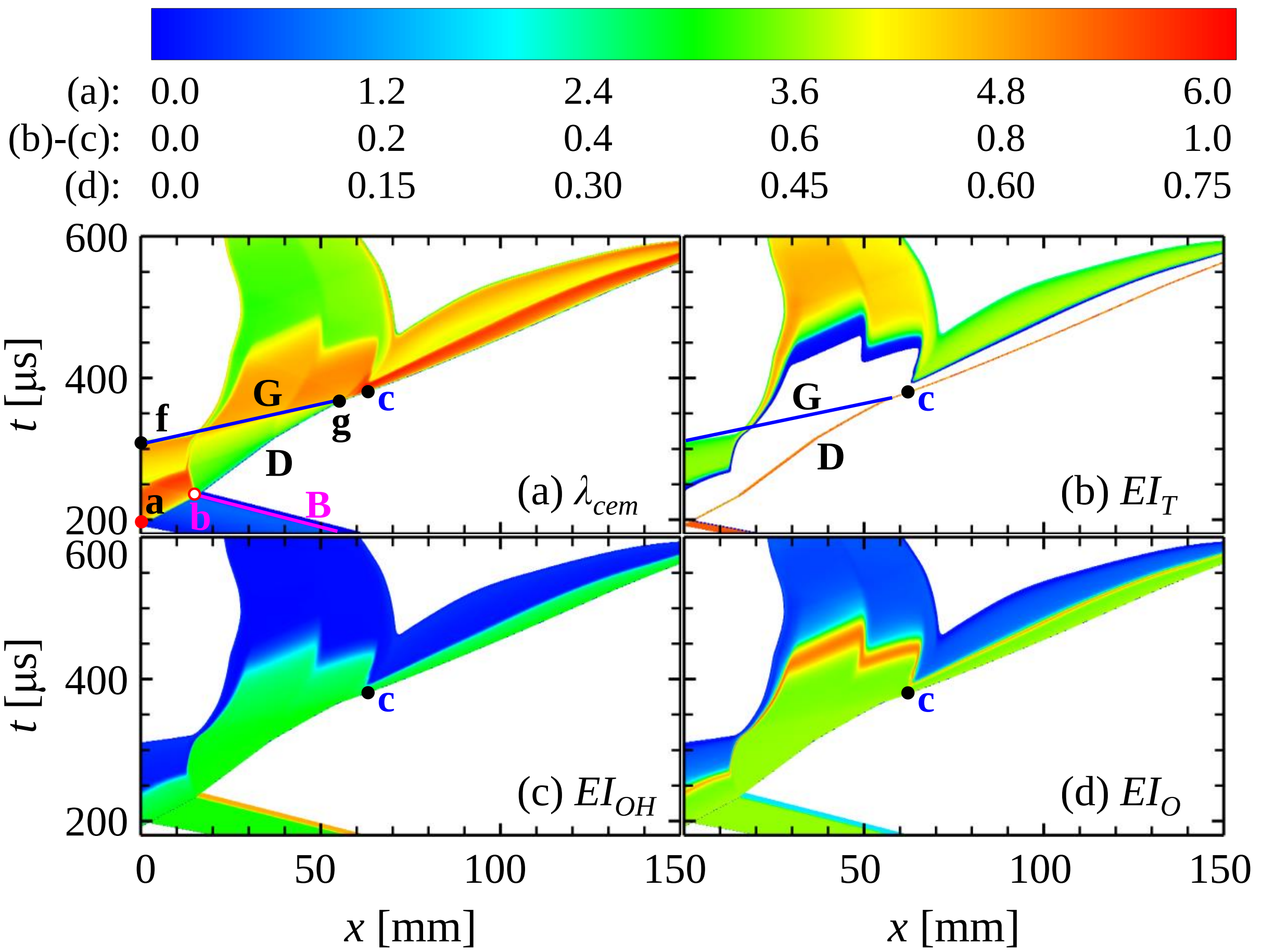

**Fig. 7.** *x-t* diagrams of (a) CEM distribution and explosion indices for (b) temperature, (c) OH and (d) O radicals.

Figure 8 further shows the evolutions of chemical timescale and OH mass fraction at $x$ = 0, 15.1 and 55.8 mm, which respectively correspond to the locations of points a, b and g in Fig. 6(a). In Fig. 8(a), the mixture is not chemically explosive (e.g., $t_e > 10^5$ s) before the incident shock wave arrives. When the incident shock reaches the three locations at $t \approx$ 192, 178 and 139 μs, respectively, $t_e$ sharply decreases to about 1 s. For the wall surface $x$ = 0, it is similar to the results in Fig. 5. For $x$ = 15.1 mm, $t_e$ slowly decreases after the incident shock wave passage because slow reactions with weak heat release are induced behind the shock (see Fig. 7). At point b, $t_e$ sharply decreases from above 0.4 s (i.e., the developing state behind the incident shock wave), to about $10^{-5}$ s (the state that is sequentially compressed by the incident shock wave and contact surface). Then $t_e$ slowly increases and decreases because of continuous chain-branching reactions, but it peaks at $t \approx$ 322 μs (a1), affected by the reaction shock wave G. After that, $t_e$ slowly increases, whereas OH radical is slowly dissociated, indicating a quasi-stable state after a deflagration flame passage. This is similar to the evolution of $t_e$ after point f at location $x = 0$.

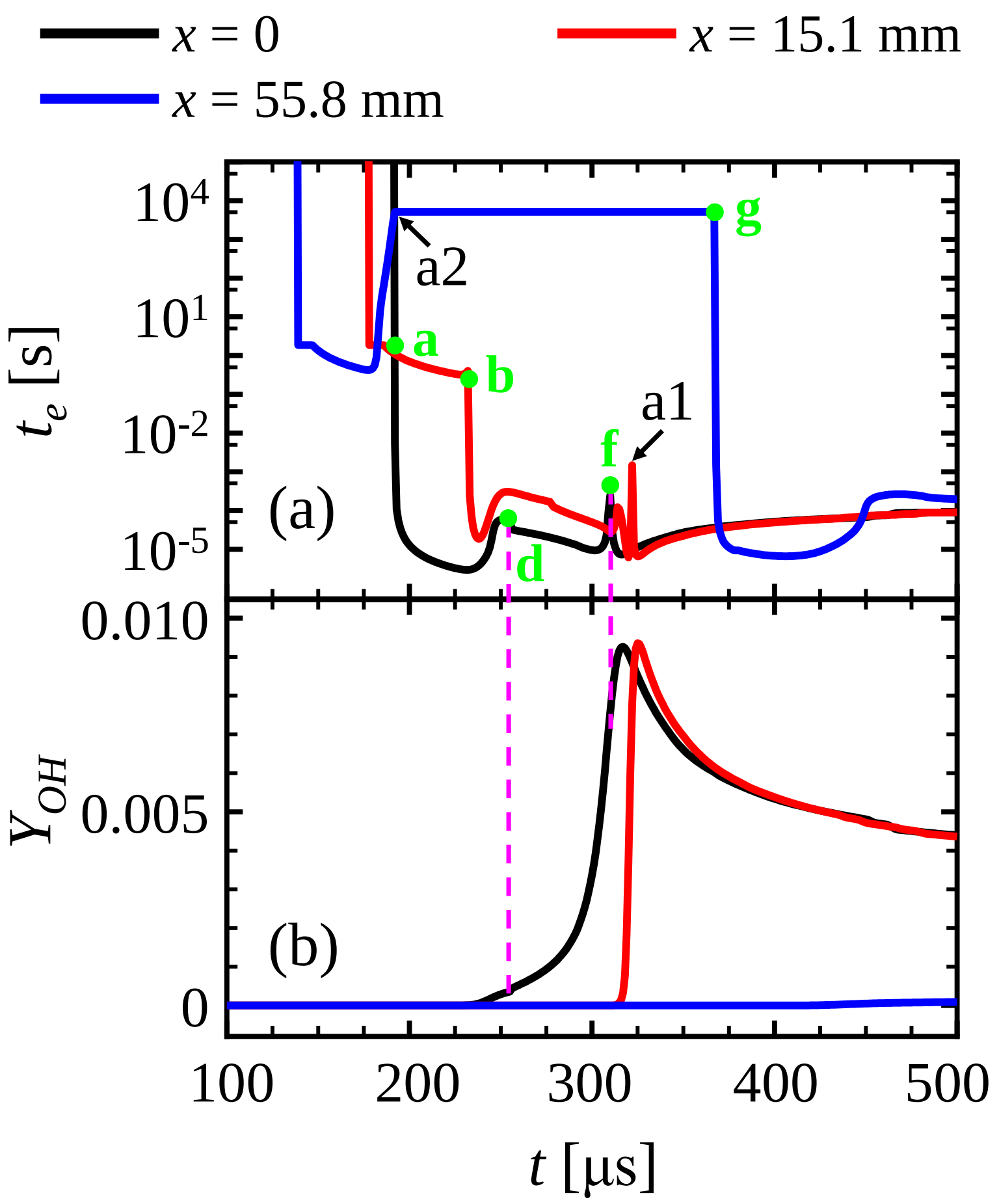

**Fig. 8.** Evolutions of (a) chemical timescale (in logarithmic scale) and (b) OH mass fraction at various locations. Letter symbols same as in Figs. 6 and 7.

For $x = 55.8$ mm, $t_e$ increases to above $10^3$ s behind the incident contact surface at $t \approx 192$ μs (a2), which ends when the reflected shock wave D and reaction shock wave G intersect here at $t \approx 366$ μs (i.e., point g). The chemical timescale drops to about $10^{-6}$ s behind g and mildly evolves as the second hot spot is slowly induced (see Fig. 6c). However, unlike the similar point g in the stoichiometric case in Fig. 5, no pronounced $Y_{OH}$ occurs till $t = 500$ μs as no detonation is developed.

*4.1.3. Deflagration behind incident shock*

Figure 9 shows the counterpart results for $M_{rw,0}$ = 3.2 and $\phi_0$ = 1.0 from mode 4. In this case, deflagration flame even occurs behind the incident shock wave, and detonation combustion is formed shortly after the reflected shock enters the combustible mixture behind the incident contact surface. In Fig. 9(a), the incident shock wave A reflects on the wall at point a ($t \approx 86$ μs). The reflected shock wave D then interacts with the incident contact surface B at point b ($x \approx 12.5$ mm and $t \approx 105$ μs) and both waves are decelerated. From points a to b, there is no obvious heat release because the mixture there is fully burned by the deflagration flame developed behind the incident shock wave A (see the high temperature between lines A and B in Fig. 9c). This differs from the lower incident shock Mach numbers in Figs. 3 and 6, where no flame occurs before the incident shock reflection. However, a deflagration flame is again immediately developed once D enters the unburned mixture behind B as seen from the heat release rate distributions in Fig. 9(d). At point e ($x \approx 15.6$ mm and $t \approx 132$ μs), the shock front is significantly intensified when it collides with the reflected contact surface F. Coupling between the reaction front and shock front occurs, leading to a developing detonation G. Although the detonation front may further interact with other waves (e.g., slightly intensified by the collision with incident rarefaction wave C at point c), it can propagate in a self-sustainable manner.

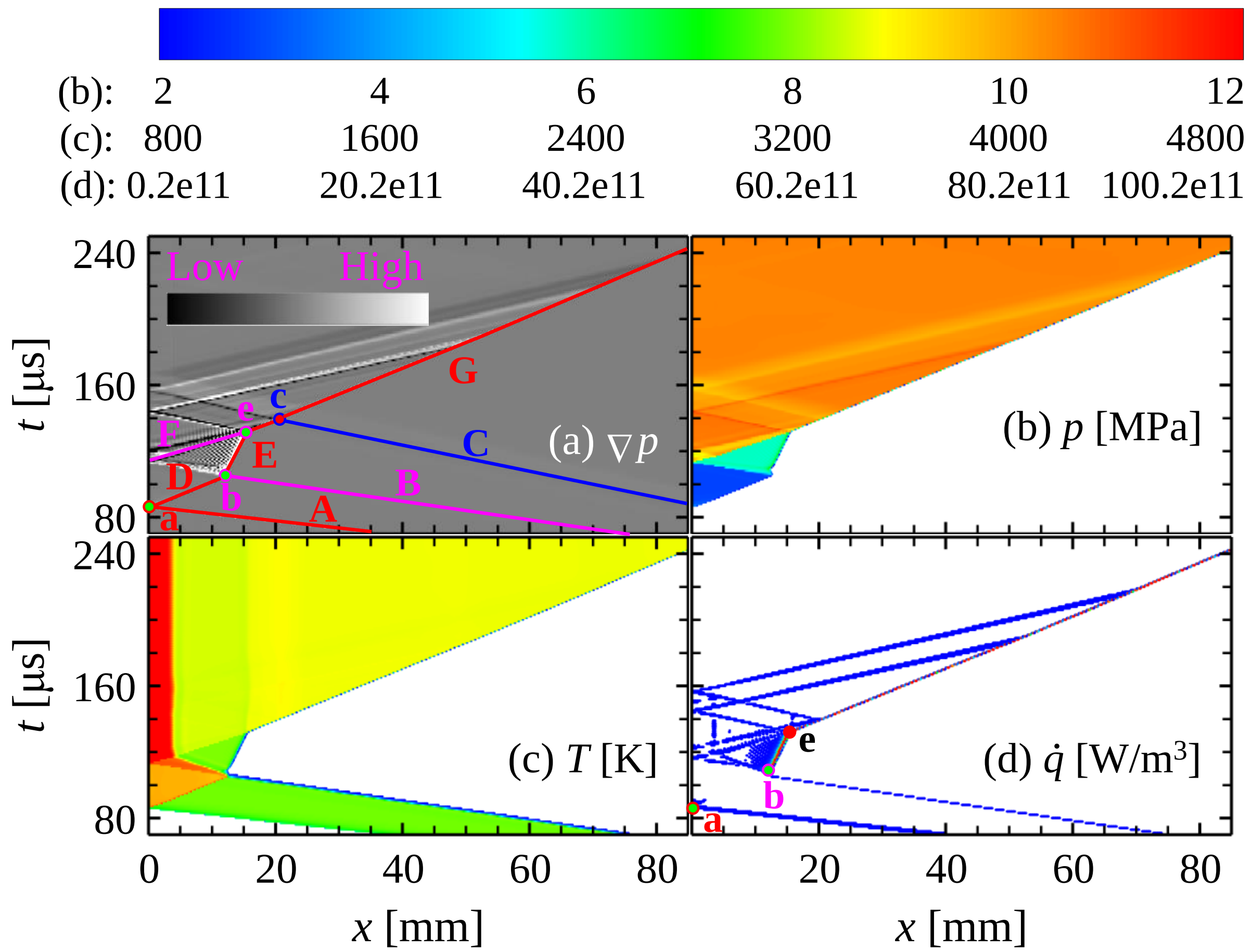

**Fig. 9.** *x-t* diagrams of (a) pressure gradient, (b) pressure, (c) temperature and (d) heat release rate. $M_{rw,0} = 3.2$ and $\phi_0 = 1.0$.

Figure 10 further shows the evolutions of chemical timescale and OH mass fraction at $x = 0$, 12.5 and 15.6 mm, which respectively correspond to the locations of points a, b and e in Fig. 9(a). In Fig. 10(a), $t_e$ is considerably reduced to about $7 \times 10^{-6}$ s when the incident shock arrives at these locations at $t \approx 86$, 81 and 80 μs, respectively. For the wall surface, $t_e$ further drops to below $5 \times 10^{-8}$ s because of the shock reflection on the wall. Meanwhile, $Y_{OH}$ also increases at point a in Fig. 10(b). After the shock reflection, $t_e$ increases to $3 \times 10^{-7}$ s and $Y_{OH}$ generally levels off over the period of interest. At $t \approx 113$ μs (arrow a1), $t_e$ drops again to about $4.5 \times 10^{-8}$ s, because the mixture is further compressed by the reflected contact surface.

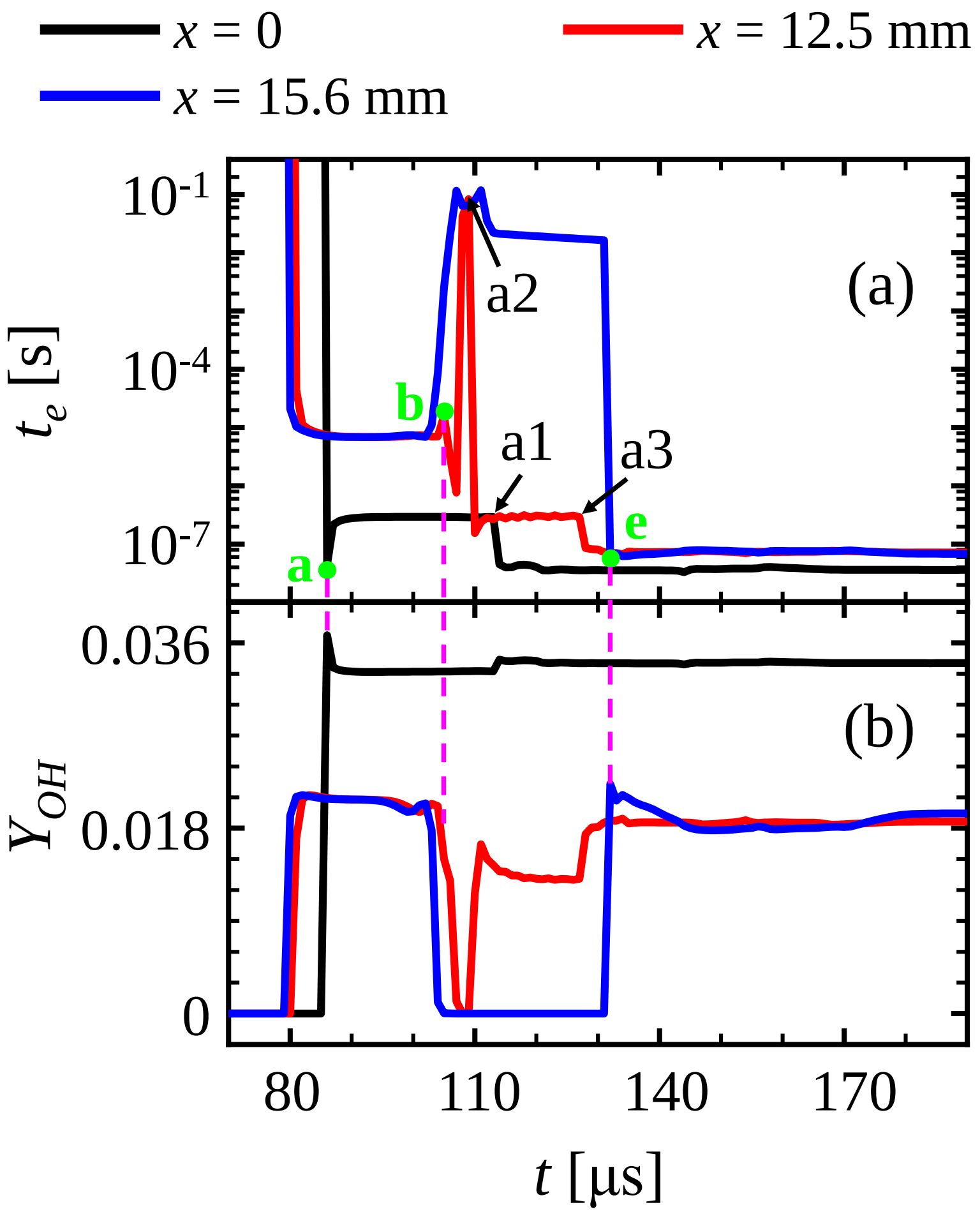

**Fig. 10.** Evolutions of (a) explosion timescale (in logarithmic scale) and (b) OH mass fraction at various locations. Letter symbols same as in Fig. 9.

For $x$ = 12.5 mm, $t_e$ is slightly increased at point b, but soon decreases as a deflagration flame is still sustained as seen in Fig. 9(d). Then $t_e$ increases to about 0.1 s at $t \approx$ 109 μs (arrow a2) because of the contact surface. Shortly, the unburned mixture is ignited by the reflected contact surface F and is fully burned at $t \approx$ 127 μs (arrow a3), both $t_e$ and $Y_{OH}$ tend to be stable. For $x$ = 15.6 mm, the unburned mixture behind the incident contact surface is directly detonation combusted by the interaction of reflected shock wave E and reflected contact surface F at point e.

*4.2. Further interpretations about wave interaction effects on chemical reaction*

In the above section, it is found that interactions between the reflected shock wave and contact surface, as well as rarefaction wave, have significant effects on hot spot formation and reaction front development. Therefore, how they affect the chemical reactions behind these unsteady events will be further investigated in this section based on one representative case in Section 4.1.1, i.e., with $M_{rw,0}$ = 2.0 and $\phi_0$ = 1.0.

Figure 11 shows the evolutions of pressure, temperature, velocity, and chemical timescale, before and after shock wave / contact surface collision (i.e., at point b in Fig. 3a, $t \approx 239$ μs). Note that across a contact surface, pressure and velocity are continuous (see Figs. 11a and 11c), whereas temperature and density are not (temperature decreases seen in Fig. 11b and density increases). The pressure ratio across a normal shock reads (in shock reference frame) [62]

$$\frac{p_2}{p_1} = \frac{2k}{k+1}(Ma_1^2 - 1) + 1, \tag{5}$$

where the subscripts '1' and '2' respectively denote the parameters before and behind the shock front, and $k$ is the specific heat ratio (here assumed to be constant for simplicity). Before the two-wave collision, the reflected shock wave propagates in the medium compressed by the incident shock wave, with $p_1$ = 104.6 kPa and $T_1$ = 751.7 K. However, after collision, they are changed to $p_1$ = 104.6 kPa and $T_1$ = 550.5 K, swept by the contact surface. Therefore, after wave collision, the reflected shock propagates into a denser but colder gas. In Fig. 11(a), it is found that the pressure ratio ($p_2/p_1$) is increased after the collision (e.g., increased from 5.21 to 5.87, from $t$ = 230 to 240 μs). Therefore, the pressure behind the reflected shock ($p_2$) is also increased as $p_1$ remains unchanged across the contact surface. According to Eq. (5), the shock Mach number, $Ma_1$, is increased after collision.

Besides, the temperature ratio across a normal shock reads [62]

$$\frac{T_2}{T_1} = \frac{[2kMa_1^2 - (k-1)][(k-1)Ma_1^2 + 2]}{(k+1)^2 Ma_1^2}. \tag{6}$$

Hence, the temperature ratio ($T_2/T_1$) is also increased with $Ma_1$ after the collision. In Fig. 11(b), $T_2/T_1$ is increased from 1.62 to 1.75 when the two waves collide. However, before the collision, $T_1$ = 751.7

K (behind the incident shock), whereas after that, $T_1$ = 550.5 K (behind the contact surface). The pronounced decrease in pre-shock gas temperature leads to significant decrease of $T_2$ after collision, as seen from Fig. 11(b). From 230 to 240 μs, the temperature behind the reflected shock wave decreases from 1218.8 K to 965.3 K. The pronounced decrease in post-shock temperature leads to the significant increase in chemical timescale, and hence after collision, the reflected shock is weakened in chemical reactivity (see the increased $t_e$ from 230 to 240 μs, right behind the shock front indicated by a3 in Fig. 11d).

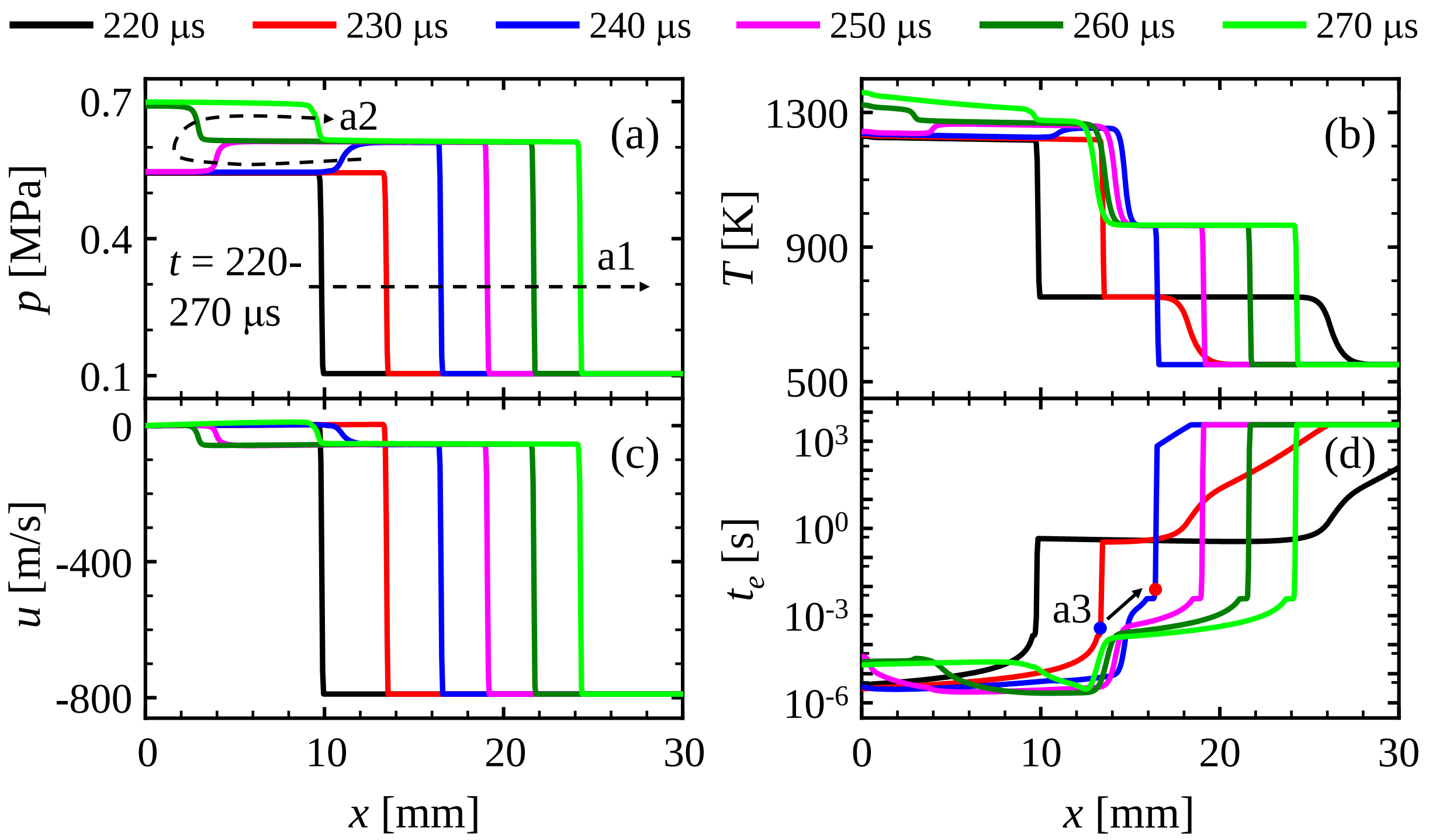

**Fig. 11.** Evolutions of (a) pressure, (b) temperature, (c) velocity and (d) chemical timescale, before ($t$ ≤ 230 μs) and after ($t$ ≥ 240 μs) the reflected shock wave / incident contact surface interaction. $M_{rw,0}$ = 2.0 and $\phi_0$ = 1.0. Arrows a1 and a2 indicate the propagation direction of reflected shock wave front and contact surface (after collision), respectively.

Figure 12 shows the evolutions of pressure, temperature, velocity, and chemical timescale, before and after the reflected shock wave / incident rarefaction wave collision (i.e., at point c in Fig. 3a, $t \approx$ 366 μs). It is well known that across a rarefaction wave, pressure, temperature, and density decrease, whereas the velocity magnitude increases. The density, pressure and temperature ratios across the reflected shock are also determined by Eqs. (5)-(6). Before interaction, the shock wave propagates in the medium that is compressed by the incident contact surface, with $p_1$ = 104.6 kPa and $T_1$ = 550.5 K. After wave collision, the reflected shock propagates in the medium behind the rarefaction wave, with $p_1$ = 77.4 kPa and $T_1$ = 509.2 K. This also indicates a decreased speed of sound in the pre-shock gas after collision. Furthermore, it is found that the propagation speed of the reflected shock increases after the collision, e.g., $u_{sf}$ = 1414.1 and 1490.6 m/s at $t$ = 360 and 370 μs, respectively. This means that the pre-shock Mach number relative to the shock front increases (i.e., $Ma_1$ ↑ because $u_1$ ↑, whereas $a_1$ ↓). Based on Eqs. (5)-(6), the pressure and temperature ratios all increase after collision. However, it does not mean that the pressure and temperature behind the shock front increase accordingly, because those in front of the shock front all decrease after the collision. Actually, the pressure and temperature behind the shock respectively change from 113.5 kPa and 1312.1 K to 101.5 kPa and 1363.5 K, from $t$ = 360 and 370 μs. The chemical reactivity behind the reflected shock increases (see the decreased $t_e$ from $t$ = 360 to 370 μs, behind the shock front indicated by a2 in Fig. 12d) with the post-shock temperature after collision. These justify why the reflected shock is intensified through colliding with the incident rarefaction wave, as observed in Section 4.1. The phenomena unveiled from Figs. 11 and 12 are also true for the reactive cases in Fig. 2.

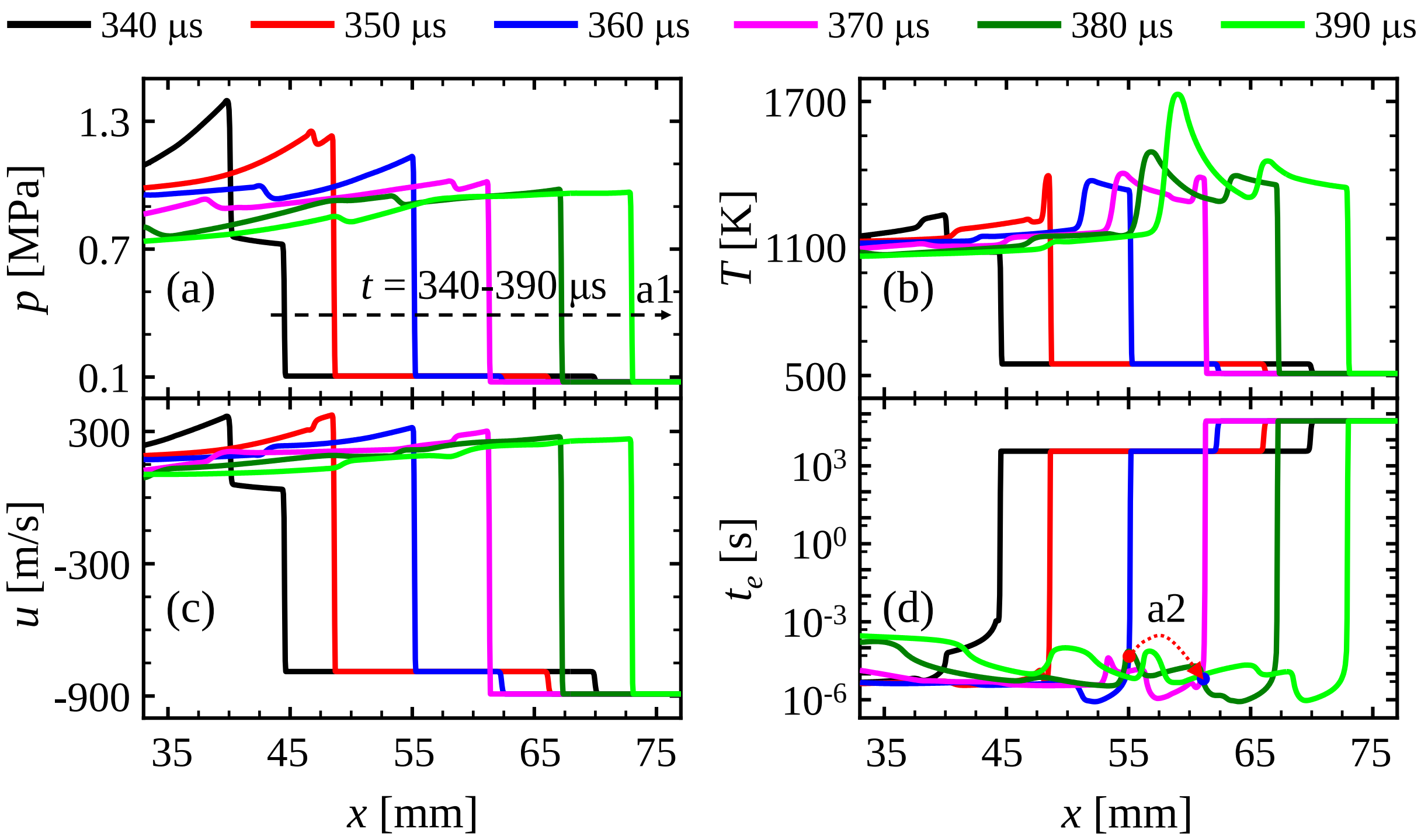

**Fig. 12.** Evolutions of (a) pressure, (b) temperature, (c) velocity and (d) chemical timescale, before ($t \leq 360$ μs) and after ($t \geq 370$ μs) the reflected shock wave / incident rarefaction wave interaction. $M_{rw,0} = 2.0$ and $\phi_0 = 1.0$.

*4.3. Multi-dimensionality effects*

Non-uniformity behind the reflected shock wave is observed both experimentally [6,63] and numerically [14–16,34,64], which is mainly caused by the interactions between the reflected shock wave and boundary layer developed behind incident shock wave. Furthermore, it is found that under "untailored" conditions, i.e., when the reflected shock interacts with a contact discontinuity, the Richtmyer−Meshkov instability can be induced [5,18,65–67]. The reflected shock is bifurcated and the uniformity (i.e., one-dimensionality) behind it is violated [6]. The multi-dimensionality effects on the combustion mode are investigated in this section.

A two-dimensional computational domain is considered. It is 0.1 m in length and 0.025 m in height ($x$ = 0-100 mm, $y$ = 0-25 mm, see Fig. 13). The 2D domain is discretized with a uniform mesh size of 10 μm, resulting in a heavy 2D calculations with 25 million cells in total. This resolution is the same as those in the 1D simulations in preceding sections. The left and right boundaries are identical to those in the 1D simulations, whereas the bottom boundary is an adiabatic non-slip wall and the top one is symmetric (see Fig. 13). The inflow Mach number is $M_{rw,0}$ = 2.4 (see Table 1), and the initial pressure and temperature of the stoichiometric mixture are also consistent with those in 1D cases, i.e., 10 kPa and 300 K.

Figure 13 shows the evolutions of pressure gradient to visualize the various wave interactions and their effects on wall boundary layers. At 70 μs, the incident shock wave A is close to the left wall, whereas the contact surface B and rarefaction wave C follow. At 80 μs, the incident shock A is reflected into A'. Meanwhile, a small separation bubble in the boundary layer occurs at the corner between the reflected shock and bottom wall (green arrow). It is induced by the reflected shock and wall boundary layer (developed behind the incident shock wave) interactions [6–8,42]. At 85 μs, the separation bubble slowly grows before the reflected shock A' collides with contact surface B. Furthermore, at 90 μs, noticeable growth of the bubble is observed after A' / B collision. At 100 μs, B evolves into B' after reflection, and A' is significantly bifurcated by the bubble. The mechanism of the reflected shock wave bifurcation is demonstrated in e.g., Refs. [6,8,16,68,69], which is mainly caused by the reflected shock / boundary layer interactions. A $\lambda$-shaped flow separation region is formed between the reflected shock and bottom wall. However, above this region, A' is generally planar in the central flow (also termed as the planar part of reflected shock [6]).

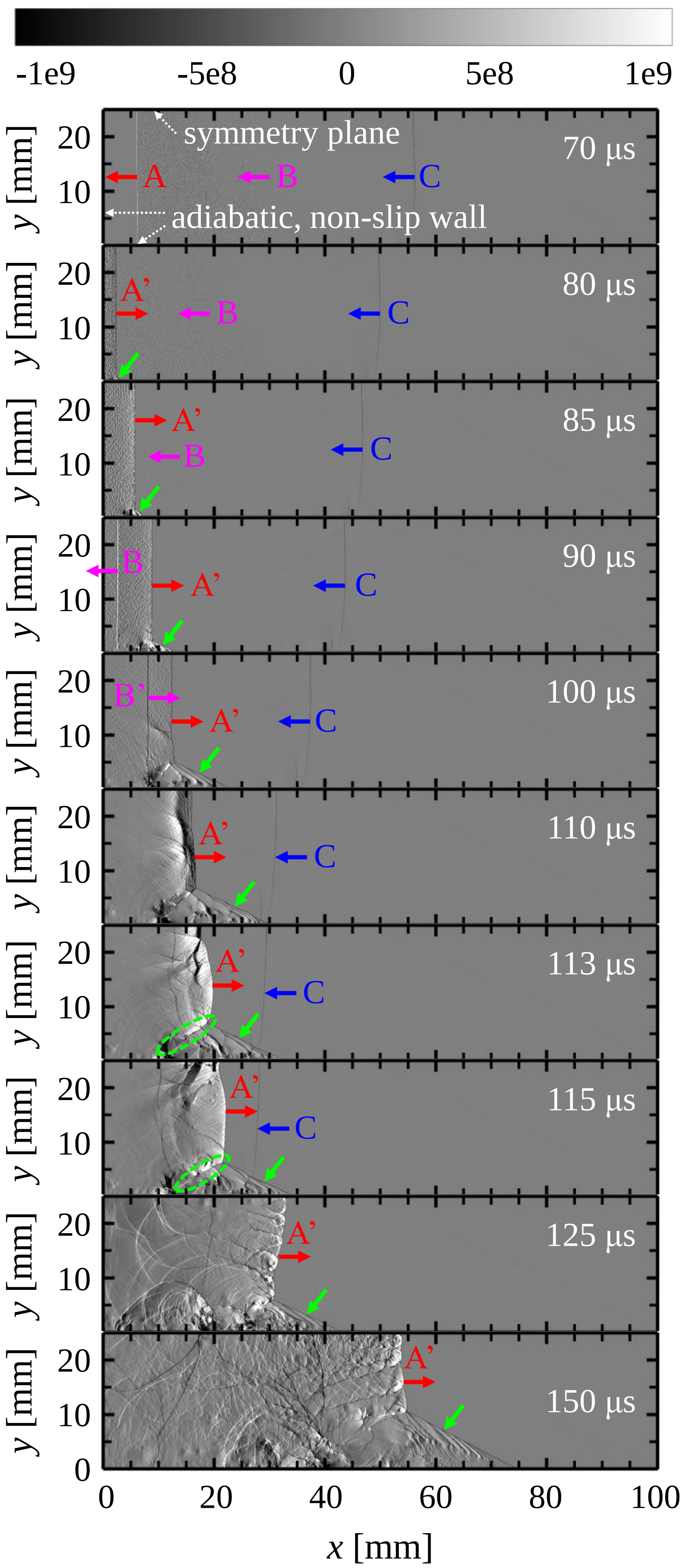

**Fig. 13.** Evolutions of pressure gradient (in Pa/m). A / A', B / B' and C respectively denote the incident / reflected shock waves, incident / reflected contact surfaces and incident rarefaction wave. $M_{rw,0}$ = 2.4 and and $\phi_0$ = 1.0.

At 110 μs, the planar part of A' is also destabilized because of its collision with reflected contact surface B'. Shortly after that, at 113 and 115 μs, the planar part of A' is distorted, i.e., above the $\lambda$-shaped region, the shock front moves slower in $x$-direction, closer to the center plane. Similar observation is also made by Yamashita *et al.* [6]. This is because the flow close to the bottom wall is compressed and hence slightly accelerated by the left bifurcation of the $\lambda$-shaped bubble (ellipses). At 125 μs, however, the shock front closer to the central region moves faster. This is because detonation combustion firstly occurs in the central region, after the collision of the reflected shock A' and incident rarefaction wave C, which will be further confirmed in Fig. 14. It is worth noting that the growth of the separation bubble is suppressed at 110-125 μs, due to the extensive heat release in the central region (see Fig. 14b). At 150 μs, the central section of the reflected shock A' is significantly compressed by the $\lambda$-shaped region due to pronounced heat release also occurs in the boundary layer, which hence expands outwards and acts as an aerodynamic throat [70].

Figures 14(a) and 14(b) respectively show the evolutions of temperature and heat release rate in the foregoing process. At 70 μs, the temperature behind the incident shock A is about 1036.8 K (see Table 1), and there is no observable heat release rate. At 80 μs, the mixture temperature is increased to 1774.7 K with obvious heat release rate (e.g., about $5 \times 10^{11}$ W/m$^3$) behind the reflected shock A'. At 85 μs, a planar deflagration flame front (with strong heat release rate of about $4 \times 10^{12}$ W/m$^3$) is developed above the separation bubble, behind the reflected shock. Therefore, the first autoignition front is purely induced from the reflected shock compression, instead of wave interactions or wall boundary layers. At 90 μs, however, the deflagration flame is significantly weakened (with heat release rate decreased to about $10^{11}$ W/m$^3$ behind the shock) because of the reflected shock wave / incident contact surface interactions. On the other hand, wall boundary layers significantly grow after the two-wave collision at 100 μs, which distort and extrude the adjacent shock, hence increase the post-shock gas temperature on the left bifurcation of the $\lambda$-shaped region (which is termed as tail shock [14,15]). Therefore, noticeable heat release rate first recovers around the shock / bubble interfaces (but still lags behind the leading shock, see the ellipse in Fig. 14b).

At 110 μs, above the separation bubble a slightly distorted deflagration flame is fully recovered after the collision of the reflected contact surface B' / shock wave A' (see Fig. 13). The flame front is close to the shock front with heat release rate above $10^{12}$ W/m$^3$. At 113 and 115 μs, the DDT process proceeds and temperature right behind the shock front is significantly increased compared with that at 110 μs. However, inside the boundary layers, temperature is pretty low with no obvious heat release, mainly because that the gas is not compressed by the reflected shock like the central flow. At 125 μs, noticeable heat release rate also occurs inside the wall boundary layers, which increases the bubble temperature and expands it outwards. The expansion of the wall boundary layers compresses the planar part of the detonation front, and the height (i.e., $y$-direction) of the latter is decreased at 150 μs. The left bifurcation of the $\lambda$-shaped region (ellipse) is also ignited by the hot burned gas. Besides, there is also increased propensity of autoignition in the right bifurcation of the bubble (arrow, which is also an oblique shock [14,15]). This is because as the bubble grows, it has increased resistance on the left-flowing gas on its right side and hence the oblique shock gets stronger. This will be further stressed in Fig. 15. Furthermore, no near-wall hot spots ever occur after 90 μs because the mixture there is already burned (see Fig. 14b). However, the near-wall temperature is still increased after 90 μs due to various wave compression.

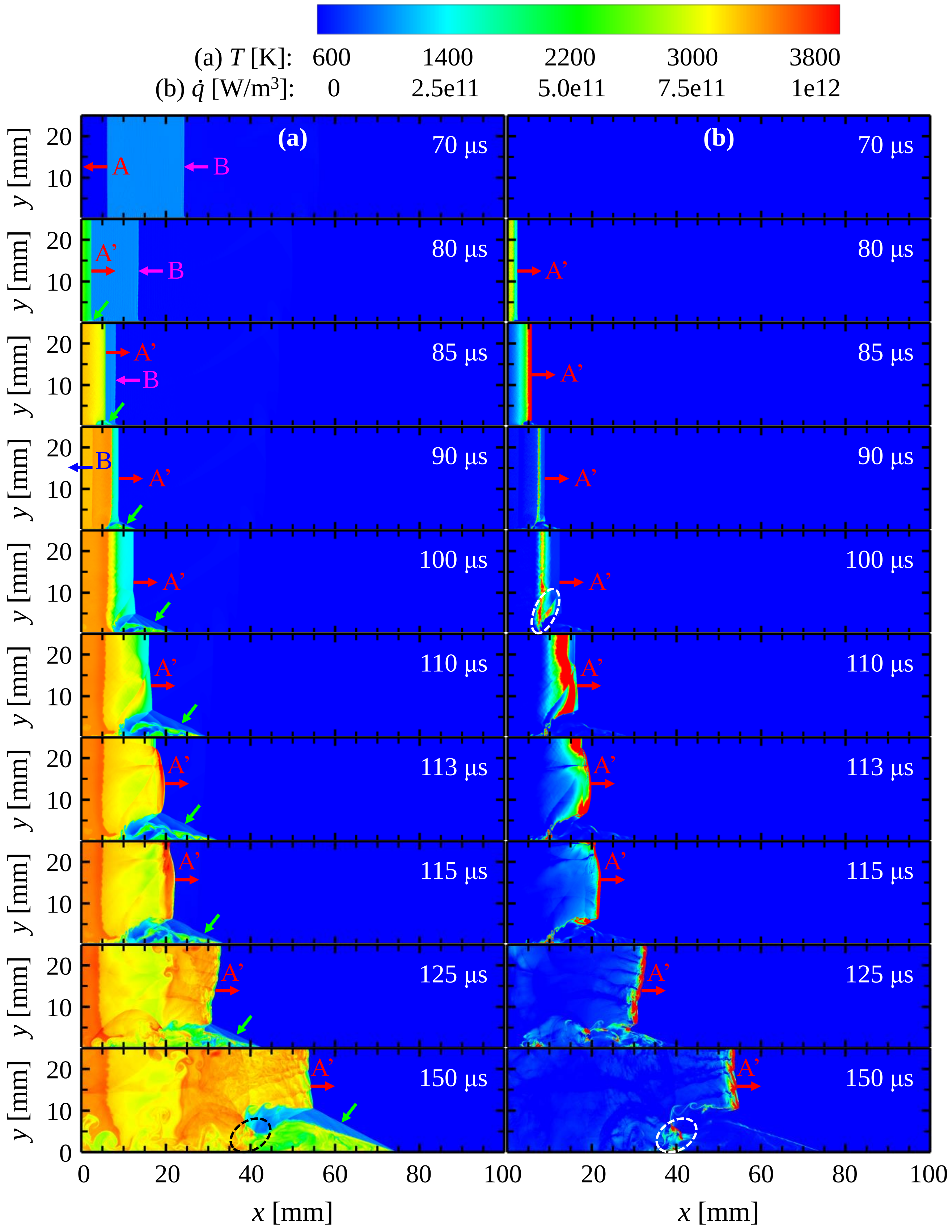

**Fig. 14.** Evolutions of (a) temperature and (b) heat release rate. Letter symbols same as in Fig. 13.

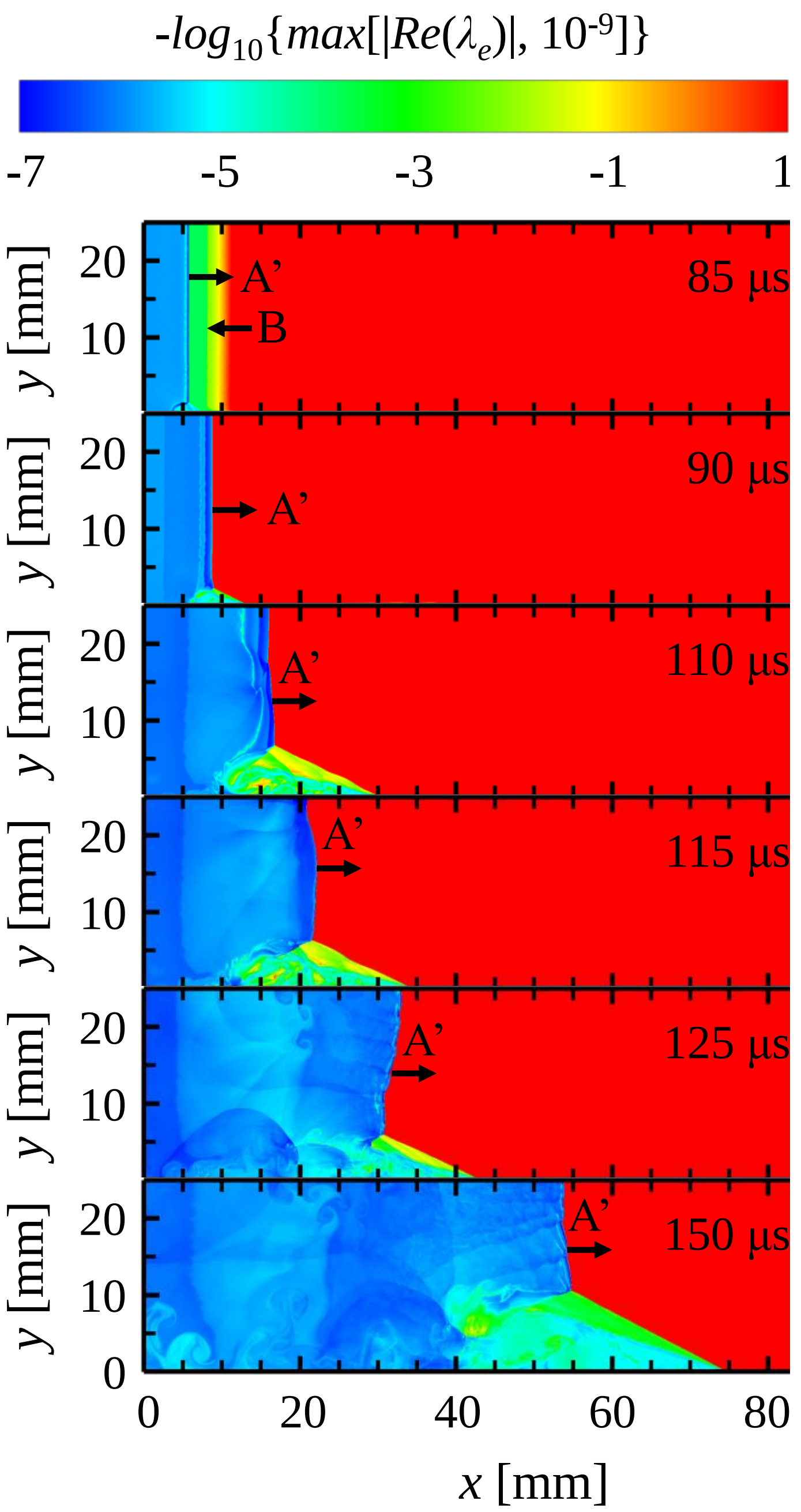

**Fig. 15.** Evolutions of chemical timescale (in logarithmic scale). Letter symbols same as in Fig. 13.

Figure 15 shows the time sequence of chemical timescale distribution to quantify the mixture reactivity affected by wave interactions and boundary layers. Behind the incident shock $t_e \approx 10^{-3}$ s, and for the un-shocked gas $t_e \gg 10$ s. Behind the reflected shock A', $t_e \approx 10^{-5}$ s. At 85-90 μs, $t_e$ further drops to about $10^{-6}$ s behind the reflected shock A' after A' / B collision, because although the deflagration flame is weakened (see Fig. 14), it is not fully quenched. At 110-115 μs, the separation bubble also shows increased reactivity. Note that the smallest $t_e$ lies between the reaction front and leading shock. However, at 125 and 150 μs, the reaction front and leading shock is fully coupled. The distribution of $t_e$ is also significantly affected by the various wave interaction in the post-shock region.

## 5. Conclusions

Autoignition and deflagration-to-detonation transition in premixed ethylene/air mixtures behind reflected shock are investigated with highly resolved numerical simulations. Skeletal mechanism for ethylene combustion is considered. Different premixture equivalence ratios ($\phi_0 = 0.2-2.0$) and incident shock Mach numbers ($M_{rw,0} = 1.8-3.2$) are studied.

A diagram describing combustion modes of ethylene/air mixture compressed by the shock is first developed. Four modes can be identified, including (1) no ignition, (2) deflagration combustion behind reflected shock, (3) detonation combustion behind reflected shock, and (4) deflagration combustion behind the incident shock (also develops to detonation behind reflected shock). Equivalence ratios and shock Mach numbers strongly affect the combustion development process. Under low $M_{rw,0}$ and/or low $\phi_0$, no ignition ($M_{rw,0} \leq 1.8$ or $M_{rw,0} = 2.0$ but $\phi_0 \leq 0.4$) or deflagration-only mode (two cases with $M_{rw,0} = 2.0$ and $\phi_0 = 0.6$, $M_{rw,0} = 2.4$ and $\phi_0 = 0.2$) is observed. Mode 3 becomes more prevalent when $M_{rw,0}$ and $\phi_0$ increases ($\phi_0 \geq 0.8$ and $M_{rw,0} = 2.0$, $\phi_0 \geq 0.4$ and $M_{rw,0} = 2.4$, $\phi_0 = 0.2$-$2.0$ and $M_{rw,0} = 2.8$). When $M_{rw,0} = 3.2$, mode 4 is observed for all the considered equivalence ratios. Moreover, the influence of equivalence ratio on combustion mode is weaker than that of inflow Mach number.

For modes 2 and 3, the gas between the incident shock and contact surface is only in the reaction induction period, whereas for mode 4 a deflagration flame is formed right behind the incident shock. Moreover, three autoignition hot spots are observed in mode 3. The first one occurs at the wall surface, induced from the sequentially re-compression of the reflected shock wave and reflected contact surface, which further develops to a reaction shock because of “the explosion in the explosion” regime. The second one is induced from the interactions between the reflected shock and incident rarefaction wave. The last one is induced by the intensified reflected shock after interacting with rarefaction wave, in the compressed mixture. It further develops to a reaction wave and couples with the reflected shock after a DDT process, and eventually detonation combustion is formed. However, in mode 2 besides the first hot spot at the wall surface, only one more hot spot is induced off the wall. It is also induced from the reflected shock / rarefaction wave collision, however, with pronounced delay. Furthermore, although it also develops to a reaction wave, it cannot catch up with the reflected shock to support a propagating detonation, before the latter arrives at the right end. For mode 4, deflagration combustion is induced by the incident shock compression whereas detonation occurs after the shock reflection.

The influence of wave interactions on chemical reactions behind the foregoing combustion modes is also interpreted. The chemical timescale from CEMA shows that the mixture reactivity decreases after the reflected shock / contact surface interaction but increases behind the incident and reflected shocks, as well as after the reflected shock / rarefaction wave interaction. Therefore, chemical reactions behind the reflected shock are weakened by contact surface, whereas intensified by rarefaction wave. The time series analysis of primitive variables including pressure, temperature and gas velocity shows that the weakening / strengthening of chemical reactions behind the reflected shock by contact surface / rarefaction wave is mainly caused by the change of pre-shock gas thermodynamic state in the reflected shock frame.

The multi-dimensionality effects are also examined with high-resolution two-dimensional simulations. The reflected shock wave / boundary layer interaction, reflected shock bifurcation, destabilization and detonation, are all observed. Bifurcation and destabilization of the reflected shock first occurs in the near wall region because of boundary layer development behind the incident shock wave. Furthermore, destabilization of the central planar part of the reflected shock is intensified by incident contact surface because of the Richtmyer – Meshkov instability mechanism. Planar autoignition is purely induced from reflected shock compression, whereas detonation combustion is formed first in the central region due to the collision of reflected shock and reflected contact surface. The left- and right-bifurcations of the separation region in the wall boundary layer are then sequentially ignited, respectively caused by the strengthened compression from the central detonation region and intensified oblique shock.

**Acknowledgements**

The computational work for this article was performed on resources of ASPIRE 1 Cluster in the National Supercomputing Center, Singapore (https://www.nscc.sg/) and Fugaku Cluster in the RIKEN Center for Computational Science in Japan (https://www.hpci-office.jp/). This work is funded by MOE Tier 1 Research Grant (R-265-000-653-114). Professor Zhuyin Ren and Dr Wantong Wu from Tsinghua University are thanked for sharing the CEMA routines.